\documentclass[twocolumn,showpacs]{revtex4}

\usepackage{graphicx}%
\usepackage{dcolumn}
\usepackage{amsmath}

\makeatletter
\def\btt#1{\texttt{\@backslashchar#1}}%
\DeclareRobustCommand\bblash{\btt{\@backslashchar}}%
\makeatother

\newcommand{\DIR}{.}

\begin{document}


\title[Short Title]{Friction effects and clogging in a 
cellular automaton model for pedestrian dynamics}

\author{Ansgar Kirchner}
\email{aki@thp.uni-koeln.de}
\affiliation{%
Institut  f\"ur Theoretische  Physik, Universit\"at zu
K\"oln D-50937 K\"oln, Germany
}%

\author{Katsuhiro Nishinari}
\email{kn@thp.uni-koeln.de}
\thanks{Permanent address: Department of Applied Mathematics and Informatics, 
Ryukoku University, Shiga, Japan (email: {\tt knishi@rins.ryukoku.ac.jp})}
\affiliation{%
Institut  f\"ur Theoretische  Physik, Universit\"at zu
K\"oln D-50937 K\"oln, Germany
}%

\author{Andreas Schadschneider}%
 \email{as@thp.uni-koeln.de}
\affiliation{%
Institut  f\"ur Theoretische  Physik, Universit\"at 
zu K\"oln D-50937 K\"oln, Germany
}%

\date{\today}

\begin{abstract}
We investigate the role of conflicts in pedestrian traffic, i.e.\
situations where two or more people try to enter the same space.
Therefore a recently introduced cellular automaton model for pedestrian 
dynamics is extended by a friction parameter $\mu$.
This parameter controls the probability that the movement of {\em all} 
particles involved in a conflict is denied at one time step. 
It is shown that these conflicts are not an undesirable artefact of
the parallel update scheme, but are important for a correct description
of the dynamics.
The friction parameter $\mu$ can be interpreted as a kind of internal 
local pressure between the pedestrians which becomes important in
regions of high density, ocurring e.g.\ in panic situations.
We present simulations of the evacuation of a large room with one door. 
It is found that friction has not only quantitative effects, but can
also lead to qualitative changes, e.g.\ of the dependence of the
evacuation time on the system parameters. We also observe
similarities to the flow of granular materials, e.g.\ arching effects.

\end{abstract}


\maketitle
\section{Introduction}
\label{intro}

Methods of physics and modern computer science have been used successfully 
for the investigation of vehicular traffic problems for a long time 
\cite{chowd,dhrev,nagatani}.
Also pedestrian dynamics has attracted some attention in recent 
years \cite{PedeEvak} and many 
interesting collective effects and self-organisation phenomena have been 
observed (for an overview see \cite{dhrev,nagatani,PedeEvak,HePED}). 

One topic which has not been studied intensively up to now (see, however,
\cite{panic}) is the
relevance of local conflicts in pedestrian traffic. A conflict indicates 
a situation in which two or more people try to enter the same space
in one timestep. 
Obviously this is a real two-dimensional effect which has no counterpart
in (directed) one-dimensional vehicular traffic. 
These conflicts are local phenomena which can have a strong influence 
on global quantities like evacuation times and flows in the presence 
of bottlenecks. Typical examples where conflicts become important are
situations with clogging and stucking encountered in crowds of panicing 
pedestrians, e.g.\ near intersections and bottlenecks. In real life this
often leads to very dangerous situations and injuries or
even fatalities during evacuations. From a physics point of view
these phenomena are related to flow properties of granular 
materials \cite{granul0,gran_1,gran_2,gran_3,granul1,granul2,granul3}.

Depending on the choice of parameters, the cellular automaton (CA) model 
introduced in \cite{ourpaper,PED01a,PED01b} is able to reproduce many of the
observed collective effects of pedestrian dynamics, i.e.\ panicing and
herding behaviour \cite{aki0}.  In this paper the model is
extended by a friction parameter $\mu$, which allows an improved
description of clogging
and stucking phenomena of pedestrians. It will be shown that conflicts 
and friction are responsible for several interesting effects which
become especially relevant in high density situations, e.g.\ during
evacuation processes, and that they are important for a correct 
reproduction of the dynamics. We start with
a short summary of the model's basic concepts and its update rules.


\section{Basic principles of the model}

\begin{figure}[h]
\begin{center}
\includegraphics[width=0.65\columnwidth]{\DIR/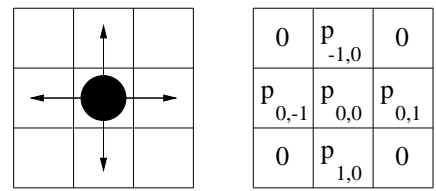}
\end{center}
\caption[]{Definition of the transition probabilities $p_{ij}$.}
\label{trans}
\end{figure}

The model considered here is a CA where the space is discretized 
into small cells which can either be empty or occupied by exactly one 
pedestrian. Each of these pedestrians can move to one of its unoccupied 
neighbour cells (see  fig.~\ref{trans}) at each discrete time step 
$t\to t+1$ according to certain transition probabilities. 
The probabilities are given by the interaction with two discrete
{\it floor fields}, $D$ and $S$ \cite{ourpaper,PED01a,PED01b,aki0}. 
The field strengths $D_{ij}$ and $S_{ij}$ at site $(i,j)$ are
interpreted as number of $D$- and $S$-particles, respectively,
present at that site.
The two fields determine the transition probability in such a way that a
particle movement in the direction of higher fields becomes more likely.

The dynamic floor field $D$ represents a virtual trace left by {\em moving}
pedestrians. Similar to the process of chemotaxis \cite{chemo,ants}
used by some insects for communication, this trace has its own dynamics,
namely diffusion and decay. Two parameters,
$\alpha$ $\in[0,1]$ and $\delta$ $\in[0,1]$, control the broadening and 
dilution of the trace. Every moving pedestrian creates a $D$-particle
at its origin cell.
Since there is no restriction of the maximal
number of $D$-particles at a site, $D$ can be regarded as a 
{\it bosonic field}. 

The static floor field $S$, on the other hand, does not change in time.
It reflects the surrounding geometry and e.g.\ specifying attractive
space regions.
In the case of the evacuation processes considered here, the static floor 
field describes the shortest distance to an exit door.
$S$ is calculated for each lattice site using some distance metric.
The field value increases in the direction of the exit such that it
is largest for door cells. An explicit construction of $S$ 
can be found in \cite{aki0,aki}.

The floor fields are used to translate a long-ranged spatial interaction
into a local interaction, but with memory, similar to the
phenomenon of chemotaxis in biology. The only other model which so
far reproduces all observed collective effects of pedestrian flow,
the social-force model \cite{social}, uses
exponentially decaying repulsive forces between pedestrians. In contrast,
in our approach the interaction is local and attractive. However,
pedestrians do not interact with the density, but with the velocity
density created by the other particles.


\subsection{Update rules}
\label{update}

The update rules of the  full model, including the interaction with 
the two floor fields, have the following structure:
\begin{enumerate}
\item The dynamic floor field $D$ is modified according to its  diffusion and
 decay rules \cite{ourpaper}, controlled by the parameters $\alpha$
and $\delta$.
In each time step of the simulation each single boson of the whole dynamic 
field $D$ decays with the probability $\delta$ and diffuses with the 
probability  $\alpha$ to one of its neighbouring cells.

\item For each pedestrian, the transition probabilities $p_{ij}$ 
for a move to an unoccupied neighbour cell $(i,j)$ (fig.~\ref{trans}) 
are determined by  the local  dynamics and the two floor fields. The 
values of the fields $D$ and $S$ are weighted with two sensitivity parameters 
  $k_S\in [0,\infty[$ and $k_D\in [0,\infty[$. This yields
\begin{equation}
\label{formula}
  p_{ij} = N\exp{\left(k_D D_{ij}\right)}
  \exp{\left(k_S S_{ij}\right)}(1-n_{ij})\xi_{ij}\,,
\end{equation}
with the occupation number $n_{ij} = 0,1$, the obstacle number
\begin{equation}
\xi_{ij} = \begin{cases} 
0 & \quad \mbox{for forbidden cells (e.g.\ walls)}\\
1 & \quad \mbox{else} \\
\end{cases}
\end{equation}
and the normalization
\begin{equation}
N = \left[\sum_{(i,j)}
e^{k_D D_{ij}} e^{k_S S_{ij}}(1-n_{ij})\xi_{ij}\right]^{-1}\,. 
\end{equation}

\item Each pedestrian chooses randomly a target cell based on the transition 
  probabilities $p_{ij}$ determined by (\ref{formula}).
\item Conflicts arising by any two or more pedestrians attempting 
to move to the same target cell are resolved by a probabilistic method.
The pedestrians which are allowed to move execute their step. The explicit 
procedure of conflict resolution will be described in detail 
in section \ref{conflicts}.
\item   $D$ at the origin cell $(i,j)$ of each {\em moving} 
particle is increased by one: $D_{ij}\to D_{ij}+1$.

\end{enumerate}
The above rules are applied to all pedestrians at the same
time (parallel update). 
This introduces a timescale of about $0.3~$sec/timestep \cite{ourpaper}.


\section{Resolution of Conflicts}
\label{conflicts}

Due to the use of parallel dynamics it is possible that two or
more particles choose the same destination cell in Step 3 of the
update procedure. Such situations will be called {\em conflicts}. 
At first it appears that conflicts are undesirable effects which 
reduce the efficiency of simulations and should therefore be avoided by 
choosing a different update scheme. We will show in the following that 
this is not the case and
that conflicts are important for a correct description of the physics
of crowd dynamics.


\subsection{Conflicts without friction}
\label{confnofric}

In \cite{ourpaper,aki0} conflicts between pedestrians were solved
in the following way: whenever $m>1$ particles
share the same target cell, one ($l\in\{1,\ldots,m\}$) is chosen to
move while its rivals for the same target keep their position. There
are two main ways to decide which particle $l$ is allowed to move:
\begin{enumerate}
\item According to the relative probabilities with which each particle 
chooses its target cell, i.e.\ the probability for particle $l$ to move 
is ${p_{ij}^{(l)}}/{\sum_{s=1}^{m}p_{ij}^{(s)}}$.
\item  All particles move with the same probability
$\frac{1}{m}$.
\end{enumerate} 
The observed behaviour has been shown to be quite robust and does not 
depend on the details of the conflict resolution  \cite{ourpaper}.


\subsection{Introduction of the friction parameter $\mu$}

\begin{figure}[h]
\begin{center}
\includegraphics[width=0.85\columnwidth]{\DIR/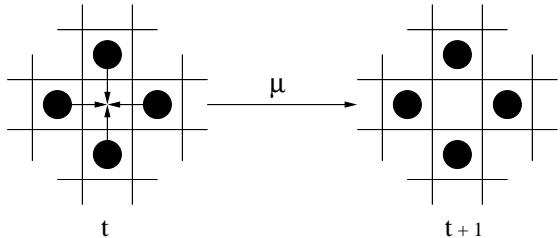}
\end{center}
\caption[]{Refused movement (for $m=4$) due to the friction parameter $\mu$. }
\label{plot_0}
\end{figure}

We now extend the basic model by a new friction parameter $\mu \in [0,1]$, 
in order to describe clogging and stucking effects between the pedestrians.
Whenever two or more pedestrians try to attempt to move to the same target 
cell, the movement of {\em all} involved particles is denied with the 
probability 
$\mu$, i.e.\ all pedestrians remain at their site (see fig.~\ref{plot_0}). 
This means that with probability $1-\mu$ one of the individuals moves to
the desired cell. Which particle actually moves is then determined
by the rules for the resolution of conflicts as described in 
Sec.~\ref{confnofric}. Note that for vanishing friction $\mu=0$
we recover the dynamics studied in \cite{ourpaper,PED01a,PED01b,aki0}.

The friction parameter might be interpreted as the effect of a
moment of hesitation: Pedestrians in conflict situations 
slow down or hesitate for a short moment when trying to resolve
the conflict. This reduces on average the velocities of all involved
particles. 

With the above definition of $\mu$ and the extended update rule it is easy
to see that $\mu$ works as some kind of local pressure between the
pedestrians. If $\mu$ is high, the pedestrians handicap each other
trying to reach their desired target sites. As we will see, this local
effect can have enormous influence on macroscopic quantities like flow
and evacuation time. We like to point out that friction as introduced
here does not reduce the velocity of a freely moving particle. Its
effects only show up in local {\em interactions}.

As one can see it is necessary for such kind of investigations to use
a parallel update in the model. Any other form of random or ordered
sequential update will disguise the real number of arising conflicts
between the pedestrians in the system.


\section{Simulations and results}

\begin{figure}[ht]
\begin{center}
\includegraphics[width=0.85\columnwidth]{\DIR/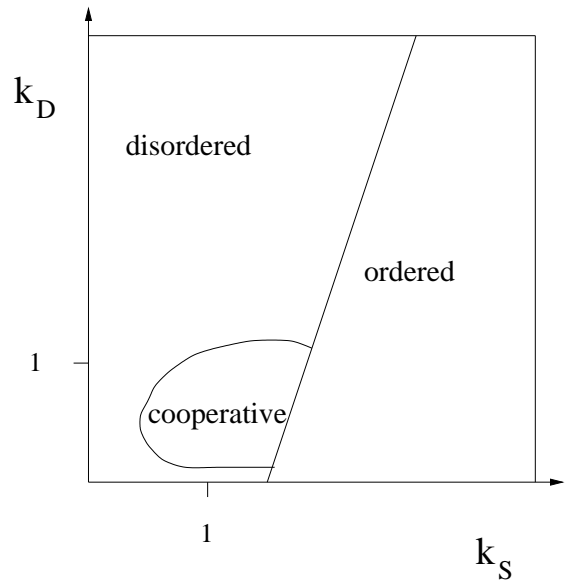}
\end{center}
\caption[]{Schematic phase diagram.}
\label{phases}
\end{figure}

In the following we describe results of simulations for a typical situation, 
i.e.\ the evacuation of a large room (e.g.\ in case of fire).
In \cite{aki0} it is shown that by variation of the sensitivity 
parameters $k_S$ and $k_D$ (see Sec.~\ref{update}) three main regimes 
for the behaviour of the particles can be 
distinguished. For strong coupling $k_S$ and very small coupling 
$k_D$ we find an {\it ordered regime} where particles only react to 
the static floor field. The behaviour then is in some sense 
deterministic.
The {\it disordered regime} characterized by strong coupling $k_D$ 
and weak coupling $k_S$ leads to a maximal value of the evacuation 
time $T$. Here the behaviour is typical for panic situations.
Between these two regimes an {\it cooperative regime} \footnote{This
regime has been denoted as `optimal regime' in \cite{aki0}.} 
exists where the combination of interactions with the static and the 
dynamic floor fields minimizes the evacuation time. 
Fig.~\ref{phases} shows a schematical phase diagram in the space of
the couplings $k_S$ and $k_D$.

Here we want to focus on the influence of the friction parameter $\mu$ on the 
evacuation times in the three main regimes. As we will see the influence of 
$\mu$ is strongest in the ordered regime. This will lead to a new 
interpretation of this regime.

\begin{figure}[ht]
\begin{center}
\includegraphics[width=0.5\columnwidth]{\DIR/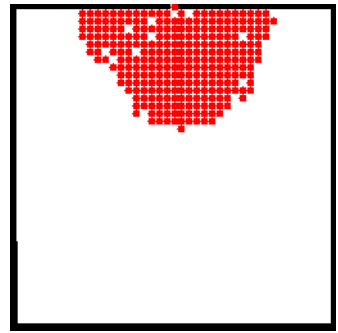}
\end{center}
\caption[]{Typical middle stage of the dynamics of an evacuation.}
\label{snap}
\end{figure}

We consider a grid of size $63\times 63$ sites with a small exit of
one cell in the middle of one wall. The particles are initially
distributed randomly and try to leave the room. The only information
they get is through the floor fields. The strength of the static floor
field at a site $(i,j)$ is inversely proportional to the distance of
$(i,j)$ from the exit measured using the metric described in \cite{aki0}. 
Fig.~\ref{snap} shows a typical
stage of the dynamics for an initial particle density of $\rho=0.3$,
i.e.\ 1116 particles.  A typical half-circle jamming
configuration in front of the door develops. 


\subsection{Influence of $\mu$ and $\rho$}

\begin{figure}[ht]
\begin{center}
\includegraphics[width=0.9\columnwidth]{\DIR/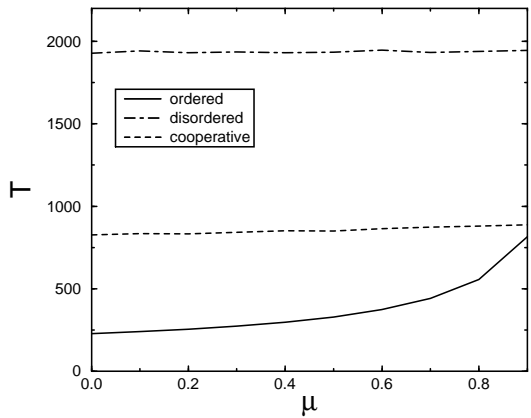}
\end{center}
\caption[]{Average evacuation times $T$ for a large room in dependence of 
the friction parameter $\mu$ for the low density $\rho=0.03$.}
\label{plot_1}
\end{figure}

\begin{figure}[ht]
\begin{center}
\includegraphics[width=0.9\columnwidth]{\DIR/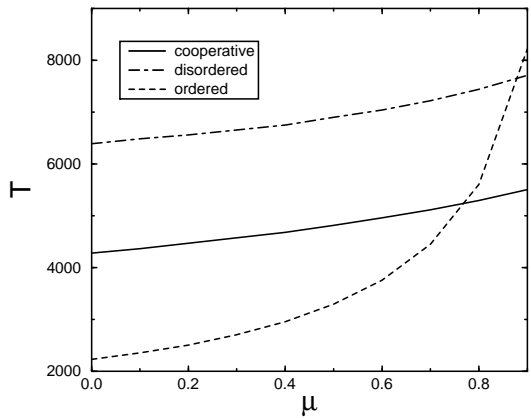}
\end{center}
\caption[]{Same as fig.~\ref{plot_1}, but for a higher density of $\rho=0.3$.}
\label{plot_1b}
\end{figure}

First we look at the averaged evacuation times $T$ in the three main
regimes, in dependence of the particle density $\rho$ and the friction
parameter $\mu$. All evacuation times are averaged over $500$ samples
and measured in update steps.  Figs.~\ref{plot_1} and \ref{plot_1b} 
show the influence of a varying $\mu$ parameter on
the three regimes for the low density $\rho=0.03$ and the high density
$\rho=0.3$.

In the low density regime, $\rho=0.03$ (fig.~\ref{plot_1}), increasing 
$\mu$ has only a very weak effect on the evacuation times of the 
disordered and the cooperative regime.  
Here the pedestrians move almost independently of each other so that
conflict situations, even close to the door, are rare since almost no
jamming occurs.
The behaviour is different in the ordered regime. All particles find 
the shortest way to the exit.
Even for low densities they will form a jam in front of the door 
after a short time.
But since these jams are small, the evacuation time 
increases only for large friction parameters $\mu>\frac{1}{2}$.

The behaviour is different in the high density regime $\rho=0.3$
(fig.~\ref{plot_1b}).  One
finds a weak increase of the evacuation time in the disordered and
cooperative regime for high $\mu$ values ($\mu > 0.6$). Even if the
particles are not packed close together in front of the door,
they form a cue and hinder each other.  Increasing $\mu$ leads to a
sharp increase of the evacuation time in the ordered regime which first
becomes larger than that in the cooperative regime and finally exceeds even
that in the disordered regime. In fact it diverges
for $\mu\longrightarrow 1$. This behaviour can be understood
from the microscopic configurations occuring.
A short time after the start of the evacuation nearly
all particles of the system are forming a big jam in front of the door. 
In this large density region many conflicts occur and for large
values of $\mu$ the outflow is strongly surpressed. Here the pressure
between the pedestrians becomes so strong that any motion is almost
impossible. Such a behaviour has been observed in panic situations
and also in simulations using the social-force model \cite{panic}. 
It will be discussed in more detail in Sec.~\ref{sec_reinter}.

\begin{figure}[ht]
\begin{center}
\includegraphics[width=0.9\columnwidth]{\DIR/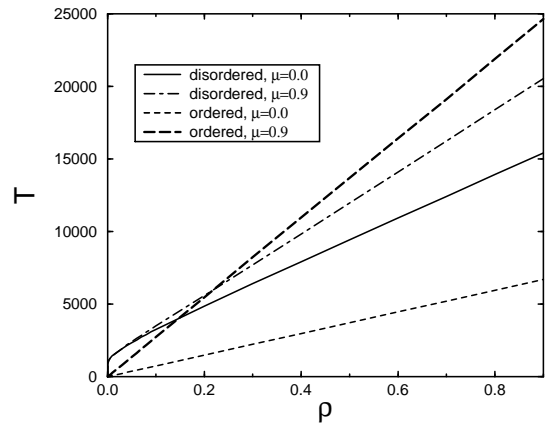}
\end{center}
\caption[]{Density-dependence of the evacuation time in the ordered 
and disordered regime for $\mu=0.9$.}
\label{plot_3}
\end{figure}

\begin{figure}[ht]
\begin{center}
\includegraphics[width=0.9\columnwidth]{\DIR/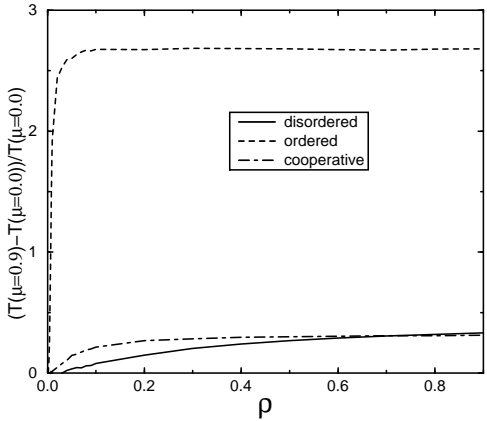}
\end{center}
\caption[]{Relative time difference of the evacuation times for
$\mu=0$ and $\mu=0.9$ in the three regimes.}
\label{plot_3b}
\end{figure}

These results are supported by figs.~\ref{plot_3} and \ref{plot_3b} 
which show the influence of an increased density $\rho$ for fixed 
values of $\mu$ on the evacuation times.
In the disordered regime even the influence of the large value $\mu =0.9$ is
only visible for higher densities ($\rho>0.1$).
The cooperative regime 
shows a similar behaviour. In contrast, in the ordered regime even for very
small densities $\rho$ the influence of the high value $\mu=0.9$ is 
very strong (fig.~\ref{plot_3}). This is supported by fig.~\ref{plot_3b}, 
where the relative time differences for the two values $\mu=0$ and 
$\mu=0.9$ for all three regimes are shown.
Here the very strong increase of the relative time difference
in the ordered regime after a very small density of $\rho=0.003$
is remarkable.
After this increase the relative time difference remains almost constant.


\subsection{Reinterpretation of the ordered phase}
\label{sec_reinter}

The introduction of the friction parameter $\mu$ allows an improved
interpretation of the ordered regime, i.e.\ the case of strong coupling 
($k_S>3$) to the static field $S$ and weak coupling to dynamic field
$D$. This regime is almost deterministic, with pedestrians
moving straight towards the exit on the shortest path.
Now an increased $\mu$ value introduces a
negative interaction between the particles into the system, i.e.\ the
pedestrians hinder each other due to strong competition for the unoccupied
target sites near the exit. So a strong coupling to $S$ together with a
high $\mu$ value, which works as an internal local pressure between
the particles, describes a typical panic situation, where an ordered
outflow is inhibited due to local conflicts near bottlenecks or doors, 
resulting in strongly increased evacuation times
(such situations are well known from emergency evacuations due to fire
or other reasons in sports arenas or passenger vessels).
\begin{figure}[ht]
\begin{center}
\includegraphics[width=0.9\columnwidth]{\DIR/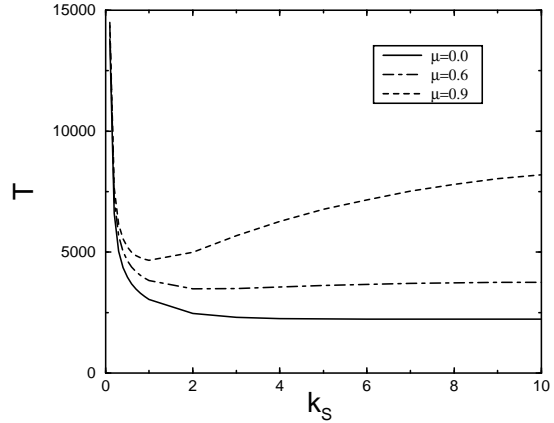}
\end{center}
\caption[]{Evacuation time as function of the sensitivity parameter 
$k_S$ for different $\mu$ values ($k_D=0$, $\rho=0.3$).}
\label{plot_4}
\end{figure}
This can be seen in fig.~\ref{plot_4}, where the influence
of an increased coupling strength to $S$ for fixed $\mu$ is
shown.  For $k_S\longrightarrow 0$ the particles 
perform a pure random walk and the evacuation times are very large
and almost independent of $\mu$. In this situation conflicts between the 
particles are not very important for the dynamics.  
In contrast, for $k_S\longrightarrow\infty$ they choose the shortest way
to the exit. Since for $\mu =0$ there is no internal pressure between 
the particles the evacuation time is minimal. However,
for $\mu\longrightarrow 1$ the number of unsolved
conflicts increases with $k_S$ due to the strong jamming at the exit. 
This results in stucking and clogging phenomena and
highly increased evacuation times. For very high $\mu$ values
($\mu=0.9$ in fig.~\ref{plot_4}) one finds a {\it minimal} evacuation
time for an intermediate coupling ($k_S\approx 1$).
This means that a larger $k_S$, which implies a larger average velocity
of freely moving pedestrians, leads to larger evacuation times.
This collective phenomenon is very similar to the faster-is-slower effect 
\cite{HePED,panic}.
Note that a similar minimum is characteristic for the cooperative
regime. However, there local minima have been observed as a function
of the coupling $k_D$ to the dynamic floor instead of $k_S$ \cite{aki0}.


\subsection{Time evolution of an evacuation}
\label{timevolv}
\begin{figure}[ht]
\begin{center}
\includegraphics[width=0.9\columnwidth]{\DIR/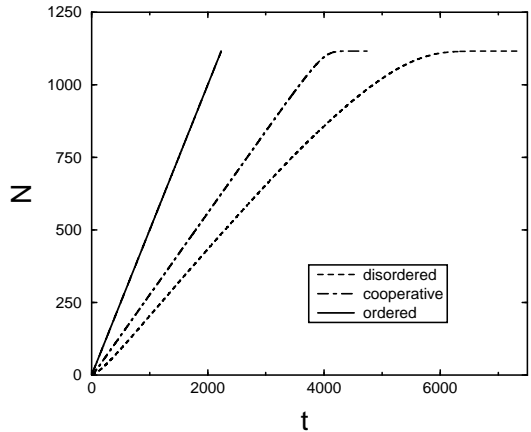}
\end{center}
\caption[]{Evacuated persons $N$ in dependence of time $t$
for all three regimes; $\rho=0.3$ and $\mu=0$.}
\label{plot_5}
\end{figure}

\begin{figure}[ht]
\begin{center}
\includegraphics[width=0.9\columnwidth]{\DIR/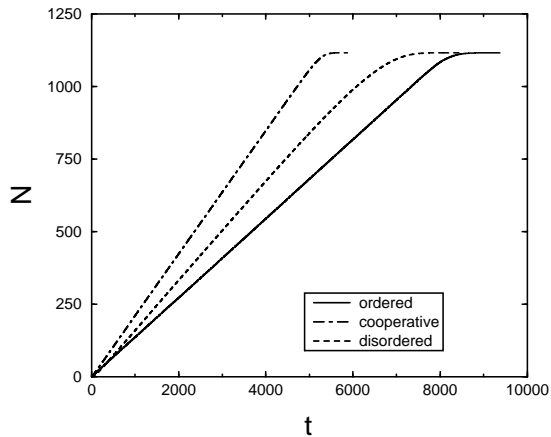}
\end{center}
\caption[]{Evacuated persons in dependence of time for all three regimes; 
$\rho=0.3$ and $\mu=0.9$.}
\label{plot_5b}
\end{figure}

In the following we look at the time evolution of an evacuation, i.e.\ 
the number of people who left the room at a certain time stage.
Figs.~\ref{plot_5} and \ref{plot_5b} show the time dependence
of the number of evacuated persons for all three regimes and different
$\mu$ values for a high density $\rho=0.3$.
The curves show a nearly linear increase since the very high initial 
density leads to strong clustering at the door already at the beginning 
of the evacuation. The evacuation times are strongly increased due to 
the large value $\mu=0.9$. This effect is again strongest in the ordered
phase. For $\mu=0$ the corresponding evacuation time is the smallest
whereas for $\mu=0.9$ it becomes the largest one of all regimes.
At the end of the evacuations, when most of the particles
have left the room, the gradient of the curves becomes rather small.

\begin{figure}[h]
\begin{center}
\includegraphics[width=0.9\columnwidth]{\DIR/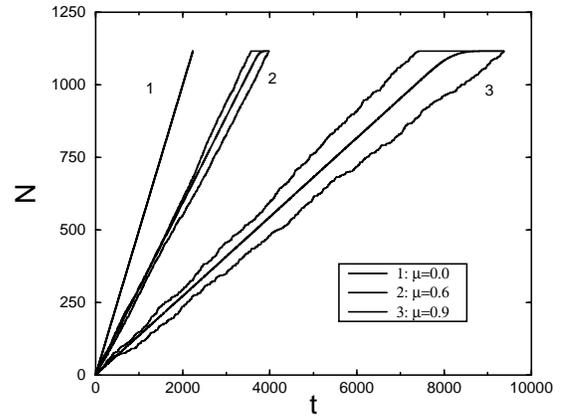}
\end{center}
\caption{Evacuated persons as function of time for the ordered 
regime for density $\rho=0.3$. Shown are the averaged, longest and 
shortest evacuation process for three $\mu$ values.}
\label{plot_6}
\end{figure}

Because of the averaging over many samples the curves are very smooth.
To have an impression of the evolution of one single evacuation and
of the variance of evacuation times, fig.~\ref{plot_6} shows the averaged
curves envelloped by the curves with the minimal and the maximal
evacuation time of the whole sample for the ordered phase and three
$\mu$ values.
For $\mu=0$ the evacuation process is nearly deterministic in the
ordered regime and the fluctuations (due to the random initial conditions 
and the dynamics) are very small.
With increasing $\mu$ values the internal pressure is increased
and the envelloping curves differ clearly from the averaged curves. 

\begin{figure}[ht]
\begin{center}
\includegraphics[width=0.9\columnwidth]{\DIR/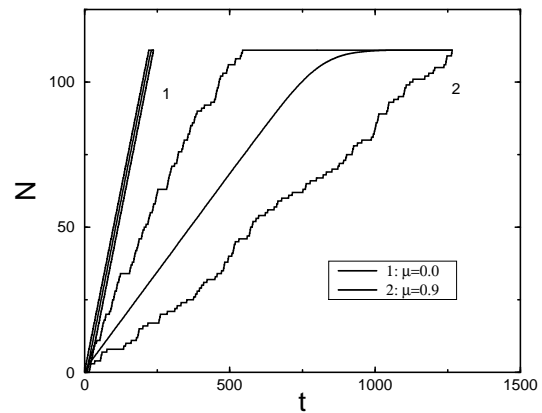}
\end{center}
\caption[]{Evacuated persons in dependence of time for the ordered 
regime; averaged, longest and shortest evacuation process for two $\mu$ 
values ($\mu=0$ and $\mu=0.9$) and density $\rho=0.03$.}
\label{plot_7}
\end{figure}

\begin{figure}[ht]
\begin{center}
\includegraphics[width=0.9\columnwidth]{\DIR/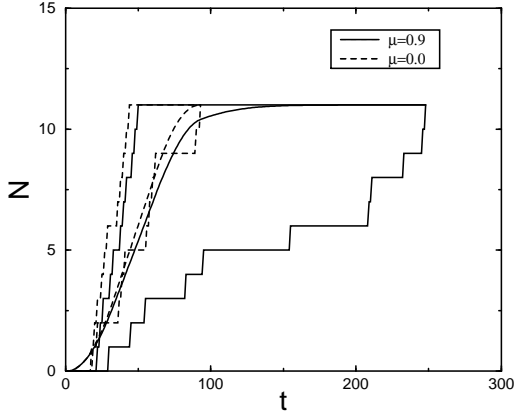}
\end{center}
\caption[]{Evacuated persons in dependence of time for the ordered 
regime; averaged, longest and shortest evacuation process for two $\mu$ 
values ($\mu=0$, $\mu=0.9$) and density $\rho=0.003$.}
\label{plot_7b}
\end{figure}

In figs.~\ref{plot_7} and \ref{plot_7b} again the averaged curves
together with the extremals are shown for the two densities $\rho=0.03$
(fig.~\ref{plot_7}) and $\rho=0.003$ (fig.~\ref{plot_7b}) and the two
friction parameters $\mu=0$ and $\mu=0.9$.
Here the same effects as for $\rho=0.3$ can be seen, but much clearer.

The time evolution of one single sample exhibits an interesting dynamics.
In figs.~\ref{plot_6}--\ref{plot_7b} small plateaus can be observed
where over short time periods no persons leave the room. This irregular 
behaviour is well-known from granular flow and is typical for clogging 
situations \cite{granul0,gran_1,gran_2,gran_3,granul1,granul2,granul3}. 
The plateaus are formed stochastically and can therefore not be observed 
after averaging over various samples (figs.~\ref{plot_5}, \ref{plot_5b}). 

Note that also the variance of the evacuation time
increases strongly with $\mu$. Fig.~\ref{plot_7} shows that the 
average evacuation time $T$ is not always a meaningful quantity for 
safety estimates since the variance can become quite large.
For the very small density $\rho=0.003$ the gradient of the curve is 
rather flat at the beginning of the evacuation (fig.~\ref{plot_7b}). 
Here there are only few particles in the system and no cue is formed
near the exit. 


\subsection{Mean-field approximation for the ordered regime}

As we have seen in Sec.~\ref{timevolv} the curve of the number of evacuated
persons $N(t)$ grows almost linearly in the ordered regime, especially for 
high densities $\rho$. For this regime we now calculate
approximatively the $\mu$-dependence of this curve, i.e.\ $N=N(t,\mu)$.

\begin{figure}[h]
\begin{center}
\includegraphics[width=0.4\columnwidth]{\DIR/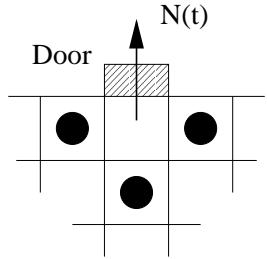}
\end{center}
\caption[]{Typical configuration in front of the door in the ordered regime.}
\label{plot_8}
\end{figure}

\begin{figure}[h]
\begin{center}
\includegraphics[width=.95\columnwidth]{\DIR/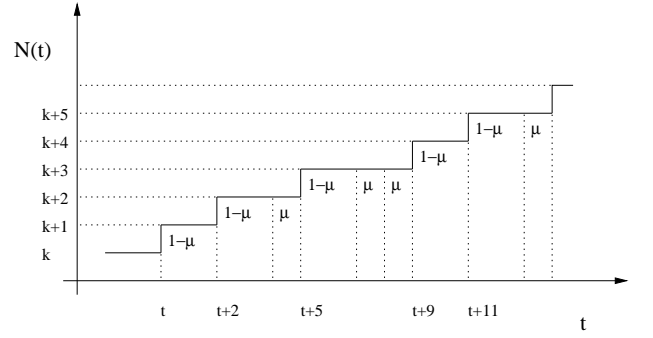}
\end{center}
\caption[]{An example for the time evolution of the number of 
evacuated persons.}
\label{plot_9}
\end{figure}

As explained earlier, in the ordered regime after a short time a big 
jam forms at the door due to the strong coupling to
the static field $S$.
Fig.~\ref{plot_8} shows a typical local configuration in front of
the exit.
Generically, 3-particle conflicts over the unoccupied lattice site in
front of the door occur. At time $t$, with probability $1-\mu$
one of these particles is able to move. In the next time
step $t+1$ this particle will escape through the door with 
probability $1$. Neglecting conflicts elsewhere in the system, because
of the big jam in front of the door the configuration shown in
fig.~\ref{plot_8} will be restored at time $t+2$.  Therefore,
repeating the above sequence a typical representation of the time
evolution of $N(t)$ is shown in fig.~\ref{plot_9}.

Using simple combinatorics, the expectation value of $N(t)$ 
can be written as 
\begin{equation}
 \langle N(t) \rangle=\frac{1}{C}\sum_{n=0}^{[t/2]}n \left(\begin{array}{c}
t-n \\
n
\end{array}\right)
 (1-\mu)^n \mu^{t-2n},
\label{mean}
\end{equation}
where $C$ is the normalization factor defined by
\begin{eqnarray}
 C(t)&=&\sum_{n=0}^{[t/2]} \left(\begin{array}{c}
t-n \\
n
\end{array}\right)
 (1-\mu)^n \mu^{t-2n}
\nonumber\\
&=&\mu^t \sum_{n=0}^{[t/2]} \left(\begin{array}{c}
t-n \\
n
\end{array}\right)
 \left(\frac{1-\mu}{\mu^2}\right)^{n}\nonumber\\
&=&\frac{1-(\mu-1)^{t+1}}{2-\mu}.\label{equat5}
\end{eqnarray}
In the last step we have used the identity (\ref{for}) which 
is derived in Appendix \ref{proof}. Due to (\ref{mean}) and 
(\ref{equat5}), $\langle N(t)\rangle$ is related to $C$ by the equation
\begin{equation}
\frac{d}{d\mu}C=\frac{t}{\mu}~C+\frac{\mu-2}{\mu(1-\mu)}~\langle 
N(t)\rangle~C
\end{equation}
or
\begin{equation}
\langle N(t)\rangle=\frac{\mu(1-\mu)}{\mu-2}\left(
\frac{d}{d\mu}\log C -\frac{t}{\mu}\right).
\end{equation}
Finally an analytical expression for $\langle N(t)\rangle$ is obtained as 
 \begin{eqnarray}
&&\langle N(t)\rangle=\frac{\mu-1}{(\mu-2)^2}\cdot\nonumber\\
&&\ \ \ \left(\mu(1+t)-2t-\frac{(\mu-2)(\mu-1)^t\mu(1+t)}{(\mu-1)^{t+1}-1}
\right).
\label{exactN}
\end{eqnarray}
Asymptotically, (\ref{exactN}) implies for large times
\begin{equation}
 \langle N(t)\rangle\sim \frac{1-\mu}{2-\mu}\cdot t
\,\,\,\quad{\rm as}\,\,\,t\to\infty\;,
\label{grad}
\end{equation}
i.e.\ a linear behaviour as observed in simulations for the
ordered regime.
This expression is also consistent with the
fact that $T\to\infty$ for $\mu\to 1$.
\begin{figure}[ht]
\begin{center}
\includegraphics[width=0.8\columnwidth]{\DIR/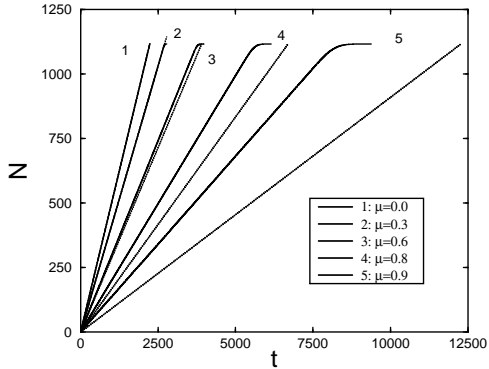}
\end{center}
\caption[]{Evacuated persons in dependence of time for the ordered 
regime; comparison of the numerical results with the analytical 
approximation (\ref{grad}) for the density $\rho=0.3$.}
\label{plot_10}
\end{figure}
\begin{figure}[ht]
\begin{center}
\includegraphics[width=0.8\columnwidth]{\DIR/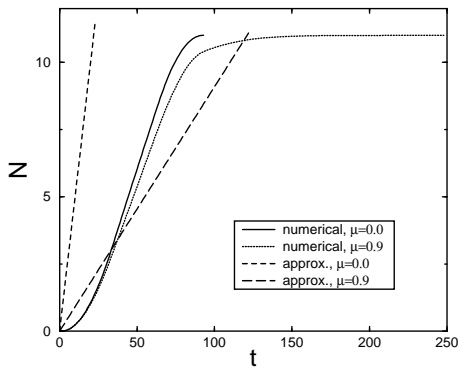}
\end{center}
\caption[]{Time-dependence of the number of evacuated persons for the 
ordered regime; comparison of the numerical results  with the analytical 
approximation (\ref{grad}) for the density $\rho=0.003$.}
\label{plot_10b}
\end{figure}
From fig.~\ref{plot_10} we see that the analytical approximation
(\ref{grad}) for $\langle N(t)\rangle$ is very good for high densities 
($\rho=0.3$) and friction parameter values $\mu\leq 0.6$. 
However, for very high parameter values ($\mu\geq 0.8$) the agreement
is not satsifactory. Here the assumption that a typical local configuration 
looks like that in fig.~\ref{plot_8} no longer holds. 
The observed flow is underestimated since the large friction parameter
$\mu$ is responsible for a flow reduction through conflicts occuring
away from the door. Therefore particles are not so dense-packed at
the exit (compared to the situation at smaller $\mu$-values) which
leads to an effective reduction of conflicts.

In the low density regime $\rho=0.03$ (not shown in the figure) 
the approximation again gives satisfactory agreement for $\mu\leq 0.6$.
Only in the region of very low densities (e.g.\ $\rho=0.003$, see 
fig.~\ref{plot_10b}) strong deviations can be observed for all $\mu$ values.
In this case there are only very few particles in the
system, not hindering each other. In most cases they reach the exit
independently and the evacuation time essentially depends on the diameter
of the room (or the average walking length) rather than 
on the number of particles in the system.


\subsection{Number of conflicts during evacuation}
\label{sec_number}

\begin{figure}[h]
\begin{center}
\includegraphics[width=0.8\columnwidth]{\DIR/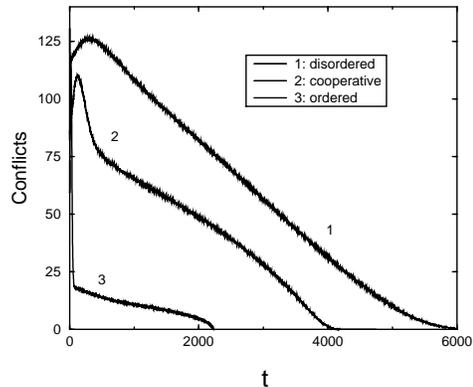}
\end{center}
\caption{Number of conflicts in dependence of time $t$ for all three 
regimes ($\rho=0.3$ and $\mu=0$).}
\label{plot_11}
\end{figure} 
\begin{figure}[h]
\begin{center}
\includegraphics[width=0.8\columnwidth]{\DIR/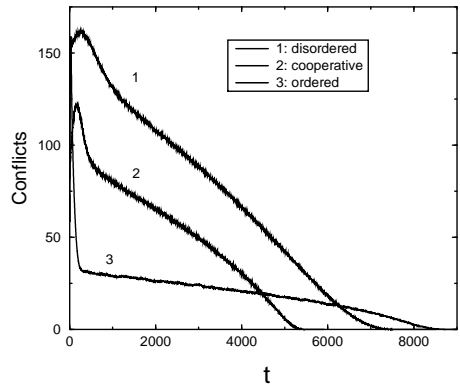}
\end{center}
\caption{Number of conflicts in dependence of time for all three 
regimes ($\rho=0.3$, $\mu=0.9$).}
\label{plot_11b}
\end{figure}
Next we investigate the time evolution of the number of conflicts arising
during an evacuation process. Figs.~\ref{plot_11} and
\ref{plot_11b} show the number of conflicts in dependence of 
time for all three regimes in the high density case $\rho=0.3$.
The maximal number of conflicts occurs for all regimes and $\mu$
values shortly after the beginning of the evacuation. The number of
conflicts is always largest in the disordered regime. 
As argued earlier here the motion resembles a random walk. 
Since there are many particles in
the system most of them will compete for empty sites with other
particles crossing their path anywhere in the room.  In the
ordered regime all pedestrians gather around the exit quickly
with most of them standing in the cue without competing
\footnote{In reality, pushing and shoving will occur. Since no motion
is possible in the jammed states of our model, this is neglected here.
It can, however, be incorporated in a more sophisticated version of
the model which e.g.\ allows to determine the pressure exerted by the
pedestrians.}.
Even though the number of conflicts is smallest in the ordered
regime, the influence of the friction parameter on the evacuation time
is strongest. The reason behind this is that nearly all conflicts
take place in front of the door. Therefore they have
a direct effect on the outflow which is reduced considerably.
Conflicts arising away from the exit do not have a direct influence
on the evacuation time. In fact they can even lead to an increased
outflow because they might surpress clogging at the door 
(see Sec.~\ref{sub_column}).

\begin{figure}[h]
\begin{center}
\includegraphics[width=0.8\columnwidth]{\DIR/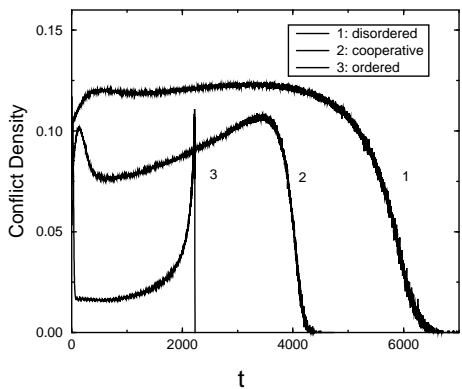}
\end{center}
\caption{Conflict density in dependence of time $t$ for all three regimes 
($\rho=0.3$ and $\mu=0$).}
\label{plot_12}
\end{figure} 
\begin{figure}[h]
\begin{center}
\includegraphics[width=0.8\columnwidth]{\DIR/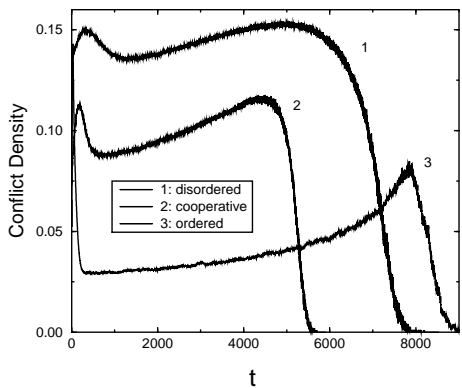}
\end{center}
\caption{Same as fig.~\ref{plot_12}, but with $\rho=0.3$ and $\mu=0.9$.}
\label{plot_12b}
\end{figure}
In figs.~\ref{plot_12} and \ref{plot_12b} the corresponding
conflict densities, i.e.\ the number of conflicts divided by the
number of particles inside the room, are shown.
In fig.~\ref{plot_12} one can see a very strong increase of the
conflict density in the ordered regime at the end of the evacuation
for $\mu=0$. During most of the time of the evacuation the majority
of particles are forming a big jam and do not contribute to
the conflict density. At the end of the evacuation the jam nearly is
dissolved and only few particles are left in the room near the exit
conflicting about the unoccupied sites.

\begin{figure}[h]
\begin{center}
\includegraphics[width=0.8\columnwidth]{\DIR/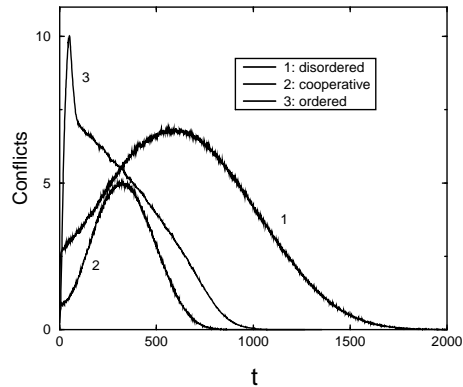}
\end{center}
\caption{Number of conflicts in dependence of the time for all three 
regimes; $\rho=0.03$ and $\mu=0$.}
\label{plot_13}
\end{figure} 
\begin{figure}[h]
\begin{center}
\includegraphics[width=0.8\columnwidth]{\DIR/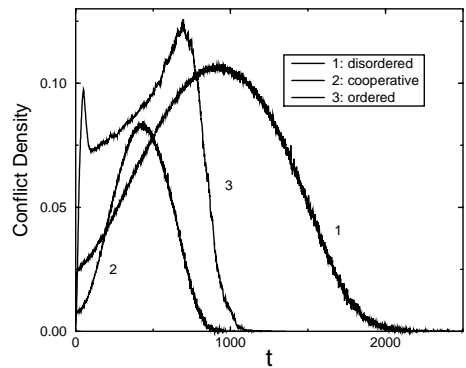}
\end{center}
\caption{Same as fig.~\ref{plot_12}, but for $\rho=0.03$ and $\mu=0.9$.}
\label{plot_14}
\end{figure}
Figs.~\ref{plot_13} and \ref{plot_14} show the number of conflicts
and the conflict density during an evacuation for the density
$\rho=0.03$ and $\mu=0.9$.
Since there are now less particles in the system,
they will not meet each other that often in the disordered and
cooperative regime. For times smaller than the evacuation time of
the ordered regime the number of conflicts and the conflict
density are largest in the ordered regime, where all particles
are packed in a small region near the exit.


\subsection{Column in front of the exit}
\label{sub_column}

As an example for safety estimations in architectural planning we 
investigate how evacuation times change if a column (i.e.\ a 
non-traversable obstacle) is placed in front of the exit.
The authors of \cite{HePED,panic}, who studied the same situation 
and found surprising results, called for complementary data
from experiments or other models to confirm their findings.
We therefore study the scenarios of \cite{HePED,panic} in the
following.

\begin{figure}[ht]
\begin{center}
\includegraphics[width=0.5\columnwidth]{\DIR/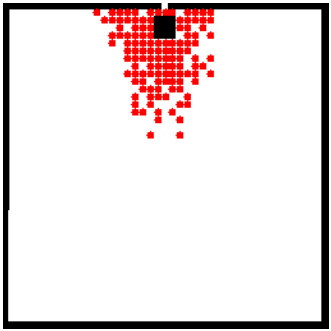}
\end{center}
\caption[]{Mid time configuration of an evacuation of a room with one door
at the middle of the top wall with an additional column of size 
$3\times 3$ cells placed in front of it. The width of the door is one 
cell.}
\label{plot_15}
\end{figure}

\begin{figure}[h]
\begin{center}
\includegraphics[width=0.8\columnwidth]{\DIR/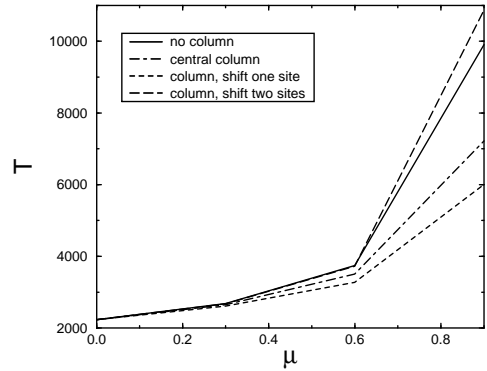}
\end{center}
\caption{Evacuation time as function of the friction parameter $\mu$ for 
four room geometries: no column, central column and column shifted by
one and two lattice sites. The initial density is $\rho=0.3$ and
the coupling strengths are $k_S=10$, $k_D=0$ corresponding to the
ordered regime.}
\label{plot_16}
\end{figure}

Fig.~\ref{plot_15} shows a mid time stage of evacuations for a column
placed central in front of the exit. The size of the column is
$3\times 3$ cells. It is placed within a distance of one cell from
the door.
We compare with situations where this column has been shifted (to the left 
or right) parallel to the exit by one or two lattice sites. 
Our focus is on the ordered regime where friction effects are strongest. 
The static floor fields $S$ used in the simulations have been
calculated using a Manhattan metric (for details see \cite{aki}).  
Fig.~\ref{plot_16} shows the averaged evacuation times 
as a function of the friction parameter $\mu$ for 
a room with different configurations of the column.
The placing of the column does not influence the evacuation times for
$\mu=0$. Here the movement is nearly deterministic and the particles
do not hinder each other. For higher values of $\mu$ the column becomes 
more and more relevant. It causes a subdivision of the crowd and decreases 
the local pressure between the particles. Therefore the evacuation times 
become smaller. They are minimized for a column placed in a slightly
asymmetric configuration in front of the exit (with a shift of one
lattice site, fig.~\ref{plot_16}). 

How can this surprising result be understood? On one hand, the column
has a certain screening effect that forces some pedestrians to take a 
detour and therefore potentially increases the evacuation time. 
On the other hand, the column subdivides the pedestrian flow and
so can lead to a reduction of conflict situations, especially
close to the exit. The competition between 
these two effects is then responsible for the nontrivial dependence
of the evacuation times on the position of the column.

\begin{figure}[h]
\begin{center}
\includegraphics[width=0.8\columnwidth]{\DIR/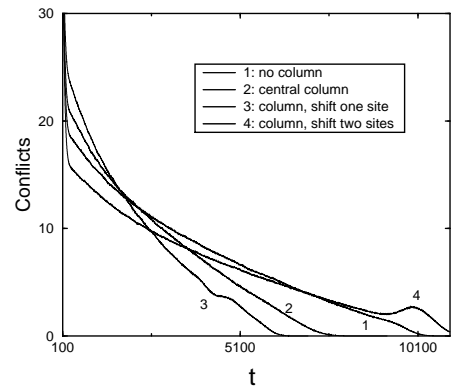}
\end{center}
\caption{Number of conflicts for a room with one door
compared to various positions of an additional column close
to the exit. The friction parameter is $\mu=0.9$. All other parameters
are the same as in fig.~\ref{plot_16}.}
\label{fig_27}
\end{figure}
Fig.~\ref{fig_27} shows the number of conflicts for three different
positions of the column in comparision to a room without column.
The evacuation time $T$ is minimal with the column shifted one site
(curve 3). In this case, the total number of conflicts during
evacuation, which is the integral of the curve, is also minimal.
However, in the early stage of the evacuation, the number 
of conflicts is bigger than for all other cases.
This is a result of the subdivision of people in the early stage of
the dynamics which occurs most pronounced in case 3.
This subdivision enables pedestrians to escape from the room faster than 
in the other cases since a smoother flow is  formed which reduces the 
evacuation time at the expense of an increasing number of conflicts 
in the early stage. Furthermore the number of dangerous conflicts
close to the door --- which have a direct influence on $T$ (see the
discussion in Sec.~\ref{sec_number}) --- is reduced.

Comparing the cases 2 and 3, the existence of the
column reduces the evacuation time in both cases due to the above
reason. However, for case~2 of a central column evacuation times
are larger than for the shifted one of case~3.
In case~2, {\em all} the people in the room have to avoid the
column and take a detour, while for a slightly shifted column
some still can reach the exit directly.
Thus a central column has a stronger screening effect which 
leads to the increase of the evacuation time compared to the slightly
shifted case.
In case~4, the column does not contribute so much to the subdivision of
the flow, but acts more like an obstacle.
For larger shifts the results will approach those of case 1.


\section{Conclusions}

We have extended a recently introduced stochastic cellular automaton 
for pedestrian dynamics by incorporating friction effects. These
are closely related to the occurance of conflicts, i.e.\ situations
where several people try to occupy the same space.
Such conflicts are only resolved with probability $1-\mu$, where
$\mu$ is called friction parameter. This implies that friction 
as introduced here does not influence the motion of a single pedestrian
(e.g.\ by reducing her average speed), but appears only through
interactions.
Friction effects become important in situations where locally high
density regions occur.  It therefore can have a strong impact
on global quantities like evacuation times although it acts only
locally.

To elucidate friction effects and the role of conflicts we have
investigated a simple evacuation scenario, especially the evacuation
time and the time evolution of the process. Without friction three different
regimes with different behaviour can be distinguished, as shown in
\cite{aki0}. Introducing $\mu >0$ affects these regimes in a different
way.

In general, the ordered regime, characterized by strong coupling to the
static field and weak coupling to the dynamic field, is affected most.
This is not surprising since for sufficiently large initial densities
large queues are formed at exits and clogging becomes relevant.
The introduction of friction leads to a strong increase of the
evacuation time (fig.~\ref{plot_1b}) due to a strong increase
of conflicts (figs.~\ref{plot_11}, \ref{plot_11b}). Here it is
important that the conflicts occur mainly very close to the exit and 
thus have an immediate influence on the outflow.
In contrast, in the other regimes conflicts typically occur everywhere
in the system and have therefore a weaker influence on the outflow
properties.

Apart form evacuation times we have also investigated the time
evolution of an evacuation process. After the introduction of friction the
outflow is irregular with periods of stucking. This leads to plateaus
in the number $\langle N(t)\rangle$ of evacuated persons after time
$t$ very similar to the behaviour observed in granular flow
\cite{granul0,gran_1,gran_2,gran_3,granul1,granul2,granul3}.  
Here the formation and breaking of arches
is responsible for the irregular flow.  Similar in our case friction
is responsible for the formation of arch-like structures that block
the flow close to the door.  The plateaus occur randomly and can
therefore not be seen after averaging over different samples. The
averaged $N(t)$-curves typically show a linear behaviour. Using simple
combinatorial arguments we found $N(t) \sim \frac{1-\mu}{2-\mu}\cdot
t$ for the $\mu$-dependence of the asymptotic behaviour of the number
of evacutated persons.  This expression agrees quite well with the
numerical data for densities not too small up to intermediate values
of $\mu$.

Finally we have shown in Sec.~\ref{sub_column} that the introduction
of friction is essential to reproduce the surprising effects observed
in \cite{panic} when placing an additional column in front of the
exit. As in \cite{panic}  this column does not necessarily 
act as an obstacle, but can --- under certain conditions, e.g.\ a slightly
asymmetric position --- improve evacuation times considerably. 
The fact that the model used here differs in many respects from the
social-force model \cite{social} (e.g.\ sign and type of interactions)
used in \cite{panic} implies a certain robustness of this phenomenon.

\begin{acknowledgments}
We like to thank  H.\ Kl\"upfel, F.\ Zielen, A.\ Kemper and D.\ Helbing
for useful discussions.
\end{acknowledgments}


\appendix
\section{}
\label{proof}

Defining the function
\begin{equation}
 a(t)=\sum_{n=0}^{[t/2]} \left(\begin{array}{c}
t-n \\
n
\end{array}\right)z^n,
\end{equation}
we have
\begin{eqnarray}
&&a(t)=
1+\sum_{n=1}^{[t/2]} \left[
\left(\begin{array}{c}
t-n-1 \\
n
\end{array}\right)+
\left(\begin{array}{c}
t-n-1 \\
n-1
\end{array}\right)\right]
z^n\nonumber\\
&&= \sum_{n=0}^{[t/2]} 
\left(\begin{array}{c}
t-n-1 \\
n
\end{array}\right)z^n 
+\sum_{n=1}^{[t/2]} 
\left(\begin{array}{c}
t-n-1 \\
n-1
\end{array}\right)
z^n.
\label{eq1}
\end{eqnarray}
The first term of the r.h.s in (\ref{eq1}) can be rewritten as
\begin{equation}
\sum_{n=0}^{[(t-1)/2]} 
\left(\begin{array}{c}
t-n-1 \\
n
\end{array}\right)
z^n=a(t-1).
\end{equation}
For odd $t$ this is obvious since then $[(t-1)/2]=[t/2]$.
For even $t$ we have $[(t-1)/2]=[t/2]-1$, but the additional
term in the sum vanishes.

The second term of the r.h.s in (\ref{eq1}) becomes
\begin{equation}
z\sum_{n=0}^{[(t-2)/2]} 
\left(\begin{array}{c}
t-n-2 \\
n
\end{array}\right)
z^n=za(t-2).
\end{equation}
Thus we obtain the recursive relation
\begin{equation}
 a(t)=a(t-1)+za(t-2).
\end{equation}
The solution of this recursion for $a(0)=a(1)=1$ is given by
\begin{equation}
a(t)=\frac{1}{\sqrt{1+4z}}\left[x_+^{t+1}-x_-^{t+1}\right]
\end{equation}
where $x_\pm$ are the roots of the quadratic equation
$x^2-x-z=0$, i.e.\ 
\begin{equation}
 x_\pm=\frac{1\pm \sqrt{1+4z}}{2}.
\end{equation}
Therefore we have
\begin{eqnarray}
&&\sum_{n=0}^{[t/2]} \left(\begin{array}{c}
t-n \\ n \end{array}\right)z^n=\nonumber\\
&&\frac{1}{\sqrt{1+4z}} \left[\left(\frac{1+\sqrt{1+4z}}{2}\right)^{t+1}
-\left(\frac{1-\sqrt{1+4z}}{2}\right)^{t+1}\right].\nonumber\\
\label{for}
\end{eqnarray}


\end{document}